\documentclass[a4paper,11pt,nohyper]{JHEP3}
\usepackage{amsmath,latexsym,amssymb}
\usepackage{graphicx}
\usepackage[latin1]{inputenc}
\usepackage{subfigure}
\usepackage{nicefrac}
\usepackage{braket}

\usepackage{comment}
\usepackage{definitions}

\title{Ricci flows and expansion in axion--dilaton cosmology}

\author{Ioannis Bakas${}^{\clubsuit}$,
Domenico~Orlando${}^{\spadesuit,\heartsuit}$ and P. Marios Petropoulos${}^{\spadesuit,\diamondsuit}$\\

  \begin{itemize}
  
  \item   Department of Physics, University of Patras \\
    26500 Patras, Greece
    
  \item   Centre de Physique Théorique, École Polytechnique\footnote{Unité
      mixte du CNRS et de l'École polytechnique, UMR 7644.} \\
    91128 Palaiseau, France
    
  \item  Theoretische Natuurkunde, Vrije Universiteit Brussel and\\
    The
    International Solvay Institutes\\
    Pleinlaan 2, 1050 Brussels, Belgium
     
  \item  Institut de Physique, Universit\'e de Neuch\^atel\\
Breguet 1, 2000 Neuch\^atel, Switzerland
  \end{itemize}

\bigskip

E-mail:
\email{bakas@ajax.physics.upatras.gr,domenico.orlando@cpht.polytechnique.fr,
marios@cpht.polytechnique.fr}}

\abstract{We study renormalization-group flows by deforming a
class of conformal sigma-models. We consider overall scale factor
perturbation  of Einstein spaces as well as more general
anisotropic deformations of three-spheres. At leading order in
$\alpha'$, renormalization-group equations turn out to be Ricci
flows. In the three-sphere background, the latter is the Halphen
system, which is exactly solvable in terms of modular forms. We
also analyze time-dependent deformations of these systems
supplemented with an extra time coordinate and time-dependent
dilaton. In some regimes time evolution is identified with
renormalization-group flow and time coordinate can appear as
Liouville field. The resulting space--time interpretation is that
of a homogeneous isotropic Friedmann--Robertson--Walker universe
in axion-dilaton cosmology. We find as general behaviour the
superposition of a big-bang (polynomial) expansion with a finite
number of oscillations at early times. Any initial anisotropy
disappears during the evolution.}

\preprint{
  CPTH-RR068.0906\\
  NEIP--06--07\\
  hep-th/0610281\\}

\begin{document}

\setcounter{footnote}{0}
\renewcommand{\thefootnote}{\arabic{footnote}}
\setcounter{section}{0}

\renewcommand{\Psi}{Q}



\section{Introduction}
\label{sec:introduction}

Finding string backgrounds with cosmological interpretation is a
challenging problem that has led to many developments, either in
genuine string theory or in string-inspired models. One possible way
to investigate string cosmology is by promoting a Euclidean
three-dimensional static string solution to a Minkowskian
four-dimensional time-dependent one. To some extent, this is borrowed
from general relativity, where \textit{e.g.} a three-sphere can be
promoted to a Friedmann--Robertson--Walker (\textsc{frw})
universe. The latter is a four-dimensional space--time which has
spherical spatial section of time-dependent radius. However, string
theory is more constraint than general relativity and such a promotion
is a delicate issue, which does not admit a systematic treatment: the
string equations of motion involve many fields (metric, dilaton and
antisymmetric tensors) and the four-dimensional universe must
necessarily be accompanied by an internal, compact space that
saturates the central charge without spoiling the perturbative (in
$\alpha'$) nature of the solution. It is not surprising therefore that
the emergence of \emph{e.g.} de Sitter solution in the context of
strings is still unclear, whereas this universe is the archetype of a
cosmological background in general relativity.

One can be more concrete and give a specific example to illustrate
the above difficulties. A three-sphere background appears in
string theory as the target space of a Wess--Zumino-Witten
(\textsc{wzw}) model. This sigma-model is an exact rational
conformal field theory. Compared to the principal model, it
contains a Wess--Zumino term, which is topological and gives rise
to a Dirac monopole quantization (put differently, the
Kalb--Ramond three-form is classified according to the third
homotopy group of the group manifold, which in the compact case is
always $\pi_3(G) = \setZ$). The radius is therefore quantized and
this creates an obstruction when trying to let it vary
continuously while keeping conformal invariance.

A possible way out would be to abandon conformal invariance. The
perturbed sigma-model at hand is no longer an infra-red fixed
point and various parameters (such as the radius of $S^3$ in the
above example) now are running with the renormalization-group
(\textsc{rg}) energy scale. It is tempting therefore, to interpret
the \textsc{rg} scale as a cosmological time and the various (new)
fixed points as steady or final states of the cosmological
evolution. Such identification is however questionable because
time is expected to appear as a genuine dimension of the target
space of a critical string and nothing a priori guarantees that
its evolution is the \textsc{rg}-flow of a non-critical string
with one dimension less. Indeed, string backgrounds satisfy
second-order equations while \textsc{rg}-flow equations are
first-order. Hence, there is no reason to trust the above approach
generically, although the dissipative nature of the dilaton may
help in some circumstances to transform second-order into
first-order equations. In some other examples, off criticality can
generate a Liouville mode that may eventually play the role of the
time direction, putting thereby on a firmer ground the
cosmological interpretation of the \textsc{rg} evolution (see for
instance Refs. \cite{David:1988hj,Distler:1988jt,Das:1990,
Sen:1990,Cooper:1991}). Despite these reservations, the method
remains useful in order to acquire a wider perspective of the
\textsc{rg} landscape around a given spatial string
solution\footnote{A renormalization-group approach to cosmology
has been put forward by other authors \cite{Carfora:1995fj}. That
perspective is however slightly different from ours.}.

Instead of abandoning criticality, one can try to keep it or
restore it. Starting from a conformal sigma-model with spatial
target space, we add an extra time direction (which requires
adjusting the internal manifold) and let some parameters
explicitly depend on it. Time evolution of these parameters is
dictated by the critical string equations of motion. This
procedure may go under the name of \emph{dynamical promotion of
the \textsc{rg}-flow} although, depending on the parameter and the
regime, this evolution may or may not be the one obtained in the
absence of time, by following the \textsc{rg}-flow. An extreme
situation correspond to the case where the parameter is a coupling
constant associated with a truly marginal deformation of the
original conformal sigma-model. Such a coupling does not run under
the \textsc{rg} whereas it can acquire in general a non-trivial
time dependence.

In the present note, we analyze both the \textsc{rg}-flow
properties and its dynamical promotion in the following framework.
We start with a conformal sigma-model with space-like
$d$-dimensional target space. We assume the metric be Einstein
with constant curvature, while criticality is achieved with a
Kalb--Ramond field and no dilaton. The off-critical excursion is
realized by switching on an overall scale factor $c$ for the
metric that does not affect the antisymmetric tensor. We perform
the \textsc{rg} analysis and show, at lowest order in $\alpha'$,
that no other fixed point exists away from $c=1$.

We further investigate the case of the three-sphere. There, in a
mini-superspace approximation, we consider a three-parameter
perturbation written in terms of left-invariant currents
describing a squashed three-sphere sustained by a Kalb--Ramond
field. The \textsc{rg}-flow equations turn out to be equivalent to
the Halphen system that we can explicitly solve in terms of
Eisenstein series. The modular properties of the latter allow to
demonstrate that, no matter the initial conditions, the system
flows towards the round sphere. Again no other fixed point exists
away from the $SU(2)$ \textsc{wzw} model.

It is remarkable that in all cases at hand -- global scale factor
or anisotropic perturbations, the \textsc{rg}-flow is in fact a
\emph{Ricci flow} driven by a connection with torsion, due to the
Kalb--Rammond background. Ricci flows have attracted much
attention recently \cite{ChowKnopf}, not only in mathematics but
also in the physics literature, in relation to various issues of
string theory such as tachyon condensation or time dependence (see
\emph{e.g.} \cite{Suyama:2005wd,Headrick:2006ti,Graham:2006gc}).

We investigate time-dependent solutions by considering a family of
backgrounds expressed as the warped product of the original
$d$-dimensional manifold with an extra time direction, while
keeping the $B$ field unperturbed\footnote{This ensures that in
the case of compact group manifolds, the quantization condition
appearing in the WZW point is satisfied despite of the warping and
is no longer an obstruction for finding new critical solutions.}.
The one-loop Weyl-invariance equations are satisfied for the full
$(d+1)$-dimensional space--time provided a background dilaton is
also switched on in the time direction. In the case of the
three-sphere, we generalize the time warping by implementing time
dependence in all three parameters that were studied in the
\textsc{rg} approach -- anisotropic warping.

The kind of solutions we obtain when only the overall scale factor
is made time-dependent (simple warping) is by construction of the
Friedmann--Robertson--Walker type. The system exhibits a friction
behaviour. Whenever the latter dominates in the solution, we
recover the \textsc{rg}-flow of the $d$-dimensional sigma-model
studied previously. In this regime, the off-critical string
generates a Liouville mode which becomes the ``missing time'' with
a background charge that is identified with the friction
coefficient. Warping coincides then with Liouville dressing. As a
consequence, the dilaton of the full $(d+1)$-dimensional target
space is linear in this regime.

In general, the solution turns out to describe a universe
undergoing a power-law expansion (big-bang solution) superimposed
with a damped oscillation governed by the dilaton and the axion.
This kind of cosmologies was considered \emph{e.g.} in
\cite{Damour:1994zq,Gasperini:2001pc}.

The case of more anisotropic warping that is studied for the
sphere can be summarized as follows: the equations of motion are
found to be a generalization of the Halphen system with a friction
term. The solution exhibits the oscillating behaviour met
previously. In particular, in the linearized limit, around the
\textsc{wzw} point, all three parameters oscillate in phase and
the solution remains qualitatively the same.

The paper is organized as follows. The \textsc{rg}-flow analysis
is performed in Sec.~\ref{sec:textscrg-flow-appr} first for the
general case with perturbation of the overall factor and then for
the three-sphere with more parameters for \textsc{rg} motion.
%
%
In Sec.~\ref{sec:equations-motion} we introduce a time coordinate,
in the manner described previously and take the various coupling
constants perturbed so far to be time-dependent. This allows us to
write down the equations of motion which are a generalization of
the equations for the \textsc{rg}-flow and take the form of the
motion of a point particle in presence of a friction term. The
system is non linear and we were not able to find a closed
solution, but both numerical and asymptotic analyzes are possible.
They show that the universe we describe grows as a superposition
of a big-bang expansion and a damped oscillation, and eventually
converges to a linear dilaton solution
\cite{Antoniadis:1988aa,Antoniadis:1988vi}. Section
\ref{sec:friedm-roberts-walk} is actually devoted to the proper
cosmological analysis of the solution, performed by moving to the
Einstein frame. A summary follows in Sec. \ref{sec:conclusions}.



\section{The renormalization-group-flow approach}
\label{sec:textscrg-flow-appr}

\subsection{The behaviour of the overall scale in general situations}
\label{sec:overall-scale}

We would like to study the \textsc{rg}-flow for the
coupling of the metric in a system without dilaton, \emph{i.e.} for the
sigma model
\begin{equation}
  \label{eq:low-dimensional-sigma}
  \mathcal{L} (c)= \frac{1}{2}  \left( c g_{\mu \nu} + B_{\mu \nu}
  \right) \partial X^\mu \bar \partial X^\nu ,
\end{equation}
where $g_{\mu \nu}$ is a $d$-dimensional Euclidean metric and $c$
and $\lambda$ are constants. The antisymmetric tensor is tuned so
that for $c = 1$ the model is conformal\footnote{One should not
confuse $c$ with the central charge.} and for simplicity we
require $g_{\mu\nu} $ to be Einstein with constant curvature (that
is always true \textit{e.g.} for group manifolds, but this case is
not exhaustive). Geometrically this means that $H = \di B$ is the
torsion that parallelizes the Riemannian metric $g_{\mu \nu}$, and
no dilaton is required for conformal invariance.

Two-dimensional sigma models are perturbatively renormalizable.
We will analyze the evolution of a
certain class of deformations with special emphasis to their
target-space interpretation, in the mini-superspace approximation.
We will perform this analysis at one loop although it is in
principle tractable at higher order.

Following
\cite{Alvarez-Gaume:1981hn,Braaten:1985is,Friedan:1980jm,Osborn:1989bu}
we define the connection
\begin{equation}
  {\Gamma^-}\ud{\mu}{\nu \rho} = \Chris{\mu}{\nu \rho} - \frac{1}{2} H\ud{\mu}{\nu \rho}
\end{equation}
and the corresponding Riemann tensor
\begin{equation}
  {R^-}\ud{\mu}{\nu \rho \sigma} = \left( 1 - \frac{1}{c^2} \right) R\ud{\mu}{ \nu \rho \sigma} .
\end{equation}
In a dimensional-regularization scheme the one-loop counterterm
reads:
\begin{eqnarray}
  \delta\mathcal{L}_{(1)} &=& \frac{\mu^\epsilon }
  {4\pi \epsilon}R^{-}_{\mu \nu} \partial X^\mu \bar \partial
  X^\nu \\
&=& \frac{\mu^\epsilon }
  {4\pi \epsilon} \frac{R}{d} \left( 1 - \frac{1 }{c^2} \right)g_{\mu \nu} \partial X^\mu \bar \partial
  X^\nu,
\end{eqnarray}
where $R$ is the Ricci scalar for $g_{\mu \nu}$. In particular, if
we start with an $SU(2)$ \textsc{wzw} model at level $k$, we have
$d=3$ and $R=6/k$.

From the above expressions, we can determine the behaviour of $c$
with respect to the scale $\mu$, captured in the corresponding
beta-function,
$\beta(c)=\nicefrac{\mathrm{d}c}{\mathrm{d}\log\mu}$. We find:
\begin{equation}
    \beta(c) = \frac{R}{2 \pi d} \left(1 -
      \frac{1}{c^2} \right),
\end{equation}
which exhibits an infra-red fixed point at $c=1$ that we already
know. It is possible to go beyond one-loop and resum higher-order
corrections. In the particular case of \textsc{wzw} models, where
the target space is a group manifold, this amounts to a  finite
renormalization of the radius, or equivalently, a shift of the
level of the affine algebra $k \to k + g^\ast$, where $g^\ast$ is
the dual Coxeter number of the algebra.

It is interesting to remark that given the form of the one-loop
counterterm, the evolution equations take the form of a
\emph{Ricci flow} where the effect of the $B$ field is taken into
account by the connection $\Gamma^-$. In fact the equations above
are equivalent to
\begin{equation}
  \frac{\di c g_{\mu \nu}}{\di \log \mu} = \frac{1}{2\pi} {R^-}_{\mu
  \nu}.
\end{equation}
As usual one should pay attention to the direction of the flow and
it is hence useful to define an \textsc{rg} time variable $\tilde
t = - \log \mu$ so that $g_{\mu \nu} (\tilde t)$ describes the
evolution of the system going towards the infra-red.

In order to compare the result with what we will find in the
following, we introduce
\begin{equation}
  \sigma ( \tilde t) = \frac{1}{2} \log c ( \tilde t) .
\end{equation}
Then the evolution of $\sigma ( \tilde t)$ going toward the
infra-red gives:
\begin{equation}
  \label{eq:energy-evol-sigma}
  \frac{\di \sigma }{\di \tilde t } =
  -\frac{ R }{4 \pi d} \mathrm{e}^{-2\sigma (\tilde t)}\left(1 - \mathrm{e}^{-4\sigma (\tilde t)}  \right)
  \equiv - \frac{1}{4\pi} V^\prime (\sigma(\tilde t))
\end{equation}
with $V(\sigma)$ given in Eq. (\ref{eq:pot}). Equation
(\ref{eq:energy-evol-sigma})
 admits the implicit solution
\begin{equation}
  \tilde t = - \frac{2 \pi d }{R} \left(  \mathrm{e}^{2\sigma(\tilde t)} -
  \arg\tanh  \left( \mathrm{e}^{2\sigma(\tilde t)} \right) \right) + \mathrm{cst.}
\end{equation}

To summarize, in $d$ dimensions and for constant-curvature
metrics, the running of the one-parameter deformation that
maintains the normalization for the topological term is a Ricci
flow, and it does not produce any new non-trivial fixed points.

\subsection{Generic perturbations of $S^3$: the Halphen system}
\label{sec:three-sphere}

The case of \textsc{wzw} is of special interest. The underlying
algebraic structure allows to switch on more general
perturbations, while it remains possible to analyze their
simultaneous evolution under the \textsc{rg} flow. Although the
issue of metric perturbation has attracted much attention in the
past (\textit{e.g.} for $O(3)$ or $O(4)$ sigma models
\cite{Fateev:1992tk,Fateev:1996ea}),  we will here investigate a
class of perturbations, for which the full analysis of the
\textsc{rg} flow can also be carried out in the presence of a
Kalb--Ramond field.

Let us for the moment consider a \textsc{wzw} model on a generic
compact semi-simple group $G$. Its action is as in Eq.
(\ref{eq:low-dimensional-sigma}) with $c=1$. The corresponding
algebra $\mathfrak{g}$ has generators $\set{T^\alpha }$ and
structure constants $f^{\alpha\beta}_{\phantom{\alpha\beta}
\gamma}$. The standard Killing form is\footnote{We will absorb the
level $k$ of the algebra in the currents. Therefore, in terms of
the structure constants, $H = f_{\alpha \beta \gamma}J^\alpha
\wedge J^\beta \wedge J^\gamma$, where the first indices in $f$
are lowered with the Killing metric ($H$ is the standard field for
the unperturbed \textsc{wzw} model). Its square is contracted with
the new metric though. Explicitly, $
  H_{\alpha \beta }^2 =
  f_{\alpha \gamma \delta} f_{\beta \gamma'
  \delta'}g^{\gamma \gamma'} g^{\delta \delta'} .
$}
\begin{equation}
  \di s^2 =  \delta_{\alpha\beta} J^\alpha \otimes J^\beta,
\end{equation}
where $J^\alpha $ are the left currents $J^\alpha = \sqrt{k} \tr
(T^\alpha g^{-1} \di g)$ for $g \in G$. We will consider a
deformed \textsc{wzw} model with metric
\begin{equation}
  \di s^2 = g_{\alpha\beta} J^\alpha \otimes J^\beta =
  \sum_\alpha \gamma_\alpha (\mu) J^\alpha \otimes J^\alpha
\end{equation}
($\mu$ is the \textsc{rg}-scale) and the standard $B$ field, which
is fixed, being a topological term. It is convenient to go to the
vielbein of the currents so that in particular the metric reads:
\begin{equation}
 \left( g_{\alpha \beta}\right) =
  \begin{pmatrix}
    \gamma_1 (\mu) \\
    & \gamma_2 (\mu) \\
    & & \ddots \\
    & & & \gamma_n (\mu)
  \end{pmatrix},
  \label{eq:defmets}
\end{equation}
where $n = \dim G$ and $\gamma_\alpha (\mu)$ arbitrary positive
functions.

The \textsc{rg}-flow equations are here
\begin{equation}
  \frac{\di g_{\alpha \beta }}{\di \log \mu} = \frac{1}{2\pi}\left(R_{\alpha \beta } -
  \frac{1}{4} H_{\alpha
  \beta}^2\right) =  \frac{1}{2 \pi}{R^{-}}_{\alpha \beta },
\end{equation}
where $R_{\alpha\beta}$ are the components of the Ricci tensor, $H
= \di B$ and $H_{\alpha \beta}^2 = H_{\alpha \gamma \delta}
H_{\beta}^{\phantom{\beta} \gamma
  \delta}$.
We observe that the \textsc{rg}-flow equations for the
perturbation pattern at hand are again governed by a Ricci flow
with torsion, as in Sec. \ref{sec:overall-scale}.

Although the full analysis is tractable for any compact group $G$,
we will here focus on the $SU(2)$, where, as we will see below,
the flow equations can be solved explicitly. The metric
(\ref{eq:defmets}) has now only three entries: $\gamma_1 (\mu),
\gamma_2 (\mu), \gamma_3 (\mu)$.
Using (\ref{eq:genric}), we obtain the following Ricci tensor:
\begin{equation}
  \left(R_{\alpha \beta} \right)= \frac{1}{2}
  \begin{pmatrix}
    \frac{\gamma_1^2 - \left( \gamma_2 - \gamma_3 \right)^2}{\gamma_2
      \gamma_3} \\
    & \frac{\gamma_2^2 - \left( \gamma_1 - \gamma_3
      \right)^2}{\gamma_1 \gamma_3} \\
    & & \frac{\gamma_3^2 - \left( \gamma_1 - \gamma_2
      \right)^2}{\gamma_1 \gamma_2}
  \end{pmatrix},
\end{equation}
while the Kalb--Ramond term reads:
\begin{equation}
  \left(H^2_{\alpha \beta}\right)= 2
  \begin{pmatrix}
    \frac{1}{\gamma_2 \gamma_3} \\
    & \frac{1}{\gamma_3 \gamma_1} \\
    & & \frac{1}{\gamma_1 \gamma_2}
  \end{pmatrix}.
\end{equation}
We introduce as before an \textsc{rg}-time pointing towards the
infra-red,
\begin{equation}
    \di \tilde t = - \frac{1}{2\pi \gamma_1(\mu) \gamma_2 (\mu) \gamma_3(\mu)} \di \log \mu ,
  \end{equation}
where we have also reabsorbed the product of the three
$\gamma_\alpha$'s. These are positive functions and such a
rescaling is therefore harmless since it does not spoil the
monotonic evolution of $\tilde{t}$ with $\mu$ ($\tilde{t}$
increases from minus infinity to plus infinity when flowing to the
infra-red from infinite to zero $\mu$). Putting everything
together one obtains the following \textsc{rg} equations:
\begin{equation}\label{eq:gamtor}
    \begin{cases}
      \displaystyle{ 2 \dfrac{\dot{\gamma}_1}{\gamma_1} = \left(\gamma_2 - \gamma_3
      \right)^2 - \gamma_1^2 + 1} ,\\[1.2em]
      2 \dfrac{\dot{\gamma}_2}{\gamma_2} = \left(\gamma_3 - \gamma_1
      \right)^2 - \gamma_2^2 + 1 , \\[1.2em]
      2 \dfrac{\dot{\gamma}_3}{\gamma_3} = \left(\gamma_1 - \gamma_2
      \right)^2 - \gamma_3^2 + 1 ,
    \end{cases}
  \end{equation}
where dot denotes the derivative with respect to $\tilde{t}$.

The evolution of Ricci flows on general homogeneous or locally
homogeneous spaces has been studied from various perspectives (see
\textit{e.g.} \cite{ChowKnopf, Isenberg:1992}). In the absence of
torsion, the last constant term in (\ref{eq:gamtor}) is missing
and the flow converges towards the round sphere, of
\emph{vanishing} radius though. The presence of torsion does not
alter this behaviour but affects the radius of the sphere which
stabilizes to $\sqrt{k}$ because all $\gamma_\alpha$'s now
converge to one\footnote{Note that the differential systems under
consideration are parabolic, hence not invariant under time
reversal.}. This non-trivial infra-red fixed point corresponds to
the $SU(2)_k$ \textsc{wzw} model.

The above results on the convergence of the flow are based on
asymptotic analysis. However, as already advertised, the
Ricci-flow equations (\ref{eq:gamtor}) can be solved explicitly in
the case at hand. Indeed, setting
\begin{equation}
  \tilde{t} =
  \log (T + T_0), \label{eq:12}
\end{equation}
which amounts in identifying $T\in [0, +\infty[$ with a monotonic
function of the inverse energy scale $\nicefrac{1}{\mu}$, and
\begin{align}
  \Omega_1 = \frac{\gamma_2 \gamma_3}{T + T_0} , &&
  \Omega_2 = \frac{\gamma_3 \gamma_1}{T + T_0} , &&
  \Omega_3 = \frac{\gamma_1 \gamma_2}{T + T_0} , \label{eq:6}
\end{align}
equations (\ref{eq:gamtor}) are recast as:
\begin{equation}
\label{eq:halphen}
  \begin{cases}
    \dot{\Omega}_1 = \Omega_2 \Omega_3 - \Omega_1 \left(\Omega_2
      + \Omega_3 \right)  , \\
    \dot{\Omega}_2 = \Omega_3 \Omega_1 - \Omega_2 \left(\Omega_3
      + \Omega_1 \right)  , \\
    \dot{\Omega}_3 = \Omega_1 \Omega_2 - \Omega_3 \left(\Omega_1 +
      \Omega_2 \right) ,
  \end{cases}
\end{equation}
where the dot now denotes the derivative with respect to the new
time\footnote{In the case of a system without torsion one would
obtain the same equations by defining $\Omega_1 = \gamma_2\,
\gamma_3$ and cyclic and keeping the same time variable $\tilde
t$.} $T$. The arbitrary constant $T_0$ does not play any
significant role.

This is the celebrated Halphen system that was studied in the 19th
century, also called Darboux--Halphen because the equations were
written by Darboux on the analysis of triply orthogonal surfaces
\cite{Darboux} and solved three years later by Halphen
\cite{halph1,halph2}. It has appeared since then in several
instances in physics\footnote{The connection to Ricci flows was
also noticed by K. Sfetsos in unpublished work.}, as \textit{e.g.}
in the search for self-dual or anti-self-dual reductions in
four-dimensional Euclidean gravity (Bianchi IX class with $SU(2)$
isometry) \cite{Gibbons:1979xn} or in the scattering of
$SU(2)$-Yang--Mills monopoles \cite{Atiyah:1985dv}. The function $
  Y= -2\left(\Omega_1 + \Omega_2 + \Omega_3 \right)
$
satisfies
\begin{equation}\label{eq:chaz}
  \dot{\ddot{Y}}=2Y\, \ddot{Y} -3\dot{Y}^2,
\end{equation}
known as the Chazy equation \cite{chaz1,chaz2}. Note that
integrable ordinary differential equations like Chazy's or
Halphen's turn out to be systematically related to self-dual
Yang--Mills reductions.

The Halphen and the Chazy equations possess remarkable properties.
If $\omega_\alpha (z)$ provide a solution to the Halphen system
for generic $z \in \mathbb{C}$ (the dot stands then for
$\nicefrac{\mathrm{d}}{\mathrm{d}z}$), then so do
  \begin{equation}
    \tilde \omega_\alpha (z) = \frac{1}{\left(c z + d  \right)^2} \omega_\alpha
    \left( \tfrac{a z + b}{c z + d} \right) + \frac{c}{c z + d} ,
    \label{eq:modtr}
  \end{equation}
with $\left(
    \begin{smallmatrix}
      a & b \\ c & d
    \end{smallmatrix}
  \right) \in PSL(2, \setC )$. Similarly, if $y(z)$ is a solution
of the Chazy equation, so is
\begin{equation}
    \tilde y (z) = \frac{1}{\left(c z + d  \right)^2} y
    \left( \tfrac{a z + b}{c z + d} \right) - \frac{6c}{c z + d}.
    \label{eq:modty}
  \end{equation}


A class of solutions are expressed as \cite{Takhtajan:1992qb}
 \begin{equation}
\label{eq:Ealpha}
    \omega_\alpha (z)= - \frac{1}{2} \frac{\di}{\di z } \log E_\alpha (z) ,
 \end{equation}
where $E_\alpha $ form a triplet of modular forms of weight two
for $\Gamma(2) \subset PSL(2,\mathbb{Z})$ (see \textit{e.g.}
\cite{ford, serre}), transforming thereby as
\begin{equation}
 z\to-\frac{1}{z}:\quad \begin{pmatrix}
   E_1  \\
   E_2 \\
   E_3
  \end{pmatrix} \to  z^2\begin{pmatrix}
   0 & 1& 0\\
   1 & 0& 0\\
   0 & 0& -1
  \end{pmatrix}
   \begin{pmatrix}
   E_1 \\
   E_2 \\
   E_3
  \end{pmatrix}\label{eq:Sgam}
\end{equation}
and
\begin{equation}
 z\to z+1 :\quad   \begin{pmatrix}
   E_1  \\
   E_2  \\
   E_3
  \end{pmatrix}\to
  -\begin{pmatrix}
   1 & 0& 0\\
   0 & 1& 0\\
   0 & 0& 1
  \end{pmatrix}  \begin{pmatrix}
   E_1  \\
   E_2  \\
   E_3
  \end{pmatrix}.
\end{equation}
The class of solutions under consideration \emph{does not} capture
situations where some $\omega_\alpha$'s are equal. In our physical
set-up, this would correspond to deformations of the three-sphere
that preserve some extra isometries, whereas all-different
$\omega_\alpha$'s restrict the isometry group to $SU(2)$,
strictly. We will discuss this possibility towards the end of the
present section, but the following should already be stressed:
Eqs. (\ref{eq:halphen}) imply that if $\Omega_\alpha=\Omega_\beta$
then $\dot{\Omega}_\alpha=\dot{\Omega}_\beta$ and therefore
$\Omega_\alpha$ and $\Omega_\beta$ never cross unless they are
equal at any time. For the moment, we will focus on the situation
where $\Omega_1\neq\Omega_2\neq\Omega_3$. We can express the
$E_\alpha$ as
\begin{equation}\label{eq:parteisenhal}
  E_1 = \frac{\nicefrac{\di \lambda}{\di z}}{\lambda},
  \quad E_2 = \frac{\nicefrac{\di \lambda}{\di z}}{\lambda -1},
   \quad E_3 =\frac{\nicefrac{\di \lambda}{\di z}}{\lambda(\lambda
   -1)},
\end{equation}
and recast the Halphen system in terms of the Schwartz equation
for $\lambda$:
\begin{equation}\label{eq:schwartz}
  \frac{\nicefrac{\di ^3\lambda}{\di z^3}}{\nicefrac{\di \lambda}{\di z}}
  -\frac{3}{2}\left(\frac{\nicefrac{\di ^2\lambda}{\di z^2}}
  {\nicefrac{\di \lambda}{\di z}}\right)^2 =
  -\frac{1}{2}\left(\frac{1}{\lambda^2}+\frac{1}{(\lambda-1)^2}-\frac{1}{\lambda(\lambda-1)}\right)
  \left(\frac{\di \lambda}{\di z}\right)^2.
\end{equation}
 The elliptic modular function
\begin{equation}\label{lamell}
  \lambda = \frac{\vartheta_2^4}{\vartheta_3^4}
\end{equation}
is a celebrated solution for this equation (see App.
\ref{sec:theta} for a reminder on Jacobi theta functions). This
provides a particular solution of the Halphen system in terms of
the three $\Gamma(2)$ weight-two \emph{Eisenstein series}.
Correspondingly,
\begin{equation}\label{eq:parteisencha}
  y = 12 \frac{\di }{\di  z } \log \eta,
\end{equation}
where $\eta$ is the Dedekind function.

For our present purposes, we must focus on real solutions
$\Omega_\alpha$ for real time $T$ ($z=iT$). These are obtained as
\begin{equation}\label{eq:omreal}
  \Omega_\alpha(T) = i  \omega_\alpha(iT) =- \frac{1}{2} \frac{\di }{\di  T } \log E_\alpha
  (iT)
\end{equation}
and
\begin{equation}
  Y(T) = i  y(iT) = \frac{\di }{\di  T } \log E_1 \, E_2 \, E_3.
\end{equation}
The modular properties of the functions under consideration set
stringent constraints between the asymptotics of the solution and
its initial conditions: large-$T$ and small-$T$ regimes are
related by $T \leftrightarrow \nicefrac{1}{T}$. Assuming
$\Omega_\alpha^0\equiv \Omega_\alpha^{\vphantom 0}(0)$
\emph{finite} and using (\ref{eq:Sgam}) and (\ref{eq:omreal}), the
asymptotic behaviour ($T \to \infty$) is found to be
\emph{universally}
\begin{equation}\label{eq:largeT}
  \Omega_\alpha = \frac{1}{T} + \mathrm{subleading}.
\end{equation}
This provides an elegant proof of the universality of generic
$SU(2)$ Ricci flows in the presence of torsion towards the
corresponding \textsc{wzw} infra-red fixed point.

\begin{figure}
  \begin{center}
    \includegraphics[width=.6\linewidth]{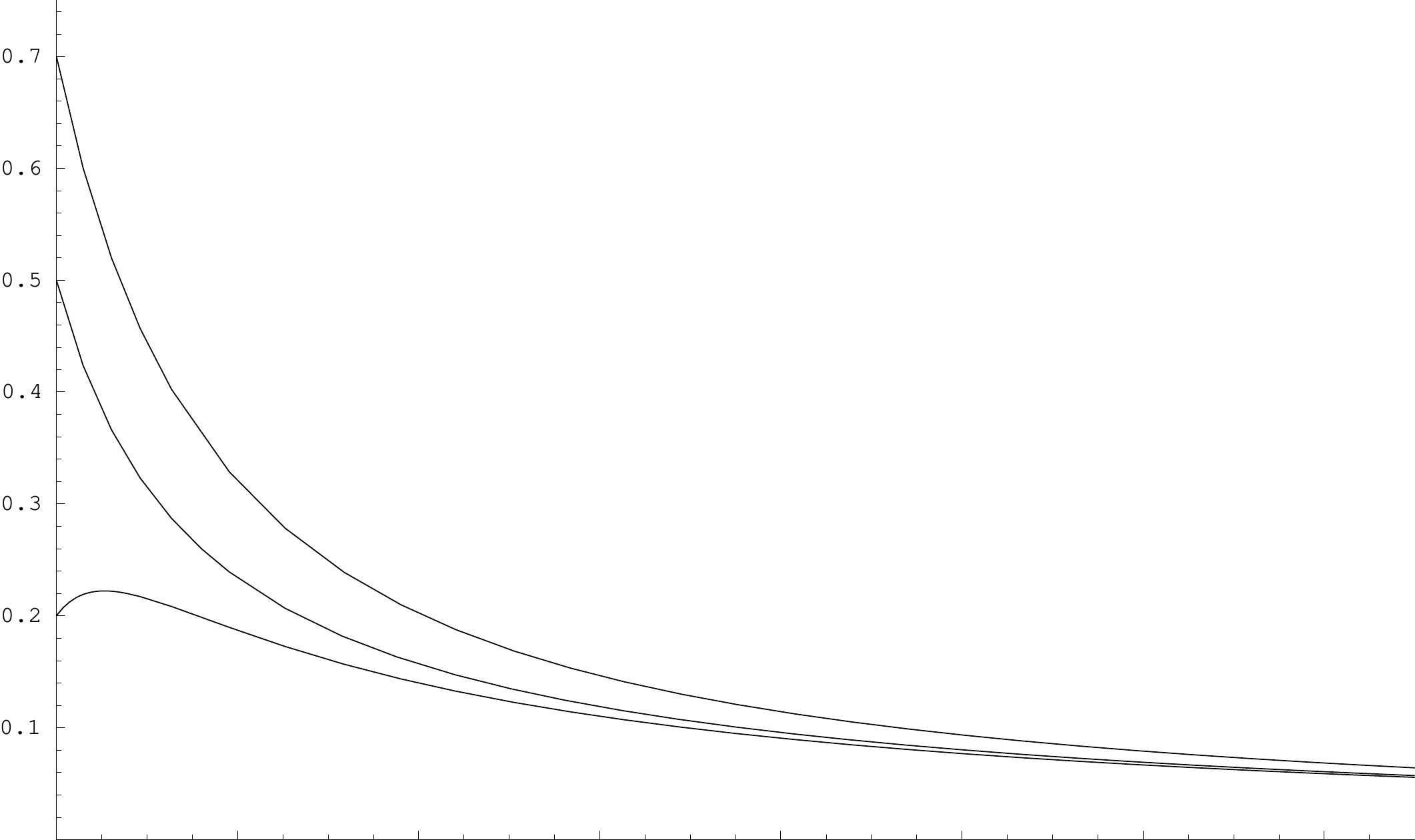}
  \end{center}
  \caption{A generic solution of the Halphen system for
positive $\Omega_1^0<\Omega_2^0<\Omega_3^0$.}
  \label{fig:gen-sol-halph}
\end{figure}

One can further assume $\Omega_\alpha^0$ \emph{finite and
positive}. This is natural for describing the initial deformation
of a three-sphere and sufficient to show that \emph{all
$\Omega_\alpha$ remain positive} at any later time. Indeed,
suppose that $0<\Omega_1^0<\Omega_2^0<\Omega_3^0$ and that
$\Omega_1$ has reached at time $t_1$ the value $\Omega_1^1=0$,
while $\Omega_2^1, \Omega_3^1>0$. From Eqs.~(\ref{eq:halphen}) we
conclude that at time $t_1$,
$\dot{\Omega}_2^1=\dot{\Omega}_3^1=-\Omega_2^1\, \Omega_3^1<0$ and
$\dot{\Omega}_1^1=\Omega_2^1\, \Omega_3^1>0$. This latter
inequality implies that $\Omega_1$ vanishes at $t_1$ while it is
increasing, passing therefore from negative to positive values.
This could only happen if $\Omega_1^0$ were negative, which
contradicts the original assumption. However, if indeed
$\Omega_1^0<0$ and $\Omega_2^0,\Omega_3^0>0$, there is a time
$t_1$ where $\Omega_1$ becomes positive and remains positive
together with $\Omega_2$ and $\Omega_3$ until they reach the
asymptotic region, where they all satisfy (\ref{eq:largeT}). The
generic behaviour of a positive-initial-value solution is given in
Fig.~\ref{fig:gen-sol-halph}.

Although the Halphen system can be solved, its deep nature makes
it difficult to establish the correspondence between a given set
of initial conditions and the modular forms $E_\alpha$ (see Eqs.
(\ref{eq:Ealpha}) or (\ref{eq:omreal})) necessary to provide the
actual solution.  Moreover, it is important to stress that
solutions exist, for which the initial conditions
$\Omega_\alpha^0$ are not all positive and not even finite. Hence,
for these solutions, (\ref{eq:largeT}) does not hold. This happens
\emph{e.g.} for the particular solution given in Eqs.
(\ref{eq:parteisenhal}) and (\ref{lamell}), where $T=0$ is a
simple pole with positive (unit) residue for $\Omega_2, \Omega_3$
and a double pole with negative residue ($-\nicefrac{\pi}{2}$) for
$\Omega_1$. As a consequence, $\Omega_1$ is negative and
increases, at large $T$, exponentially towards zero, while
$\Omega_3$ is positive and decreases exponentially towards zero;
$\Omega_2$ is positive and decreases exponentially towards
$\nicefrac{\pi}{2}$. Solutions with negative $\Omega$'s were
considered in Refs. \cite{Atiyah:1985dv,Gibbons:1986df} for the
description of the configuration manifold of two $SU(2)$
monopoles.

The existence of poles is generic for \emph{all} solutions of the
Halphen system. For  solutions corresponding to a set of finite
and different initial values, these poles are pushed behind $T=0$.
This is related to the following general property: Halphen's
solutions possess a natural movable boundary
\cite{Takhtajan:1992qb}. They exist in a domain of $\mathbb{C}$,
where they are holomorphic and single valued, and this domain has
a boundary that contains a dense set of essential singularities.
The precise location of this boundary depends on the initial
conditions.

We would like finally to mention that the Halphen system admits
also solutions which are \emph{not} based on elliptic functions.
These solutions have power-like dependence on time:
\begin{equation}
  \label{eq:polsol}
    \begin{cases}
     \Omega_1=\frac{1}{T+A}+\frac{C}{(T+A)^2},
       \\
     \Omega_2=\Omega_3=\frac{1}{T+A}.
    \end{cases}
  \end{equation}
They are actually the \emph{most general} solutions with
$\Omega_2=\Omega_3$ and describe axisymmetric deformations of the
three-sphere\footnote{Notice that the equations at hand being
first order, setting $\Omega_2^0=\Omega_3^0$ guarantees that
$\Omega_2=\Omega_3$ at any subsequent time.}, namely deformations
preserving an $SU(2) \times U(1)$ . The constants $(A,C)$ are
arbitrary and determined by the initial conditions. This class is
closed under $PSL(2, \setC )$ transformations. Indeed, using
(\ref{eq:modtr}), we learn that under $T\to \nicefrac{1}{T}$,
$(A,C)\to (\nicefrac{1}{A},\nicefrac{-C}{A^2})$, while for $T\to
T+1$, $(A,C)\to(A+1,C)$. Regularity requires $\Omega_{2,3}^0\equiv
\nicefrac{1}{A}>0$, which ensures that the pole of $\Omega_\alpha$
is located at $T<0$. Furthermore, the asymptotic behaviour is
again universal as in Eq. (\ref{eq:largeT}). No new fixed point
appears therefore in this case either. The \textsc{rg} flow of
perturbations of this type, namely with $\gamma_1 \neq \gamma_2=
\gamma_3$, was analyzed in \cite{Fateev:1996ea}, without torsion
though. When translated into the language of self-dual
four-dimensional Euclidean metrics with $SU(2)$ isometry,
solutions (\ref{eq:polsol}) correspond to the general Taub--NUT
family, including Eguchi--Hanson metrics \cite{Gibbons:1979xn}.

The case of $\Omega_1=\Omega_2=\Omega_3$ corresponds to fully
isotropic deformations studied in the previous section. In this
case, the solution is unique: $\Omega_\alpha=\nicefrac{1}{T+A}$.

To summarize and conclude the present analysis, solutions of the
Halphen system fall in three classes:
$\Omega_1\neq\Omega_2\neq\Omega_3$,
$\Omega_1\neq\Omega_2=\Omega_3$ and $\Omega_1=\Omega_2=\Omega_3$.
The corresponding three-manifolds, target spaces of the
sigma-model, are homogeneous with isometry group $SU(2)$, $SU(2)
\times U(1)$ and $SU(2)\times SU(2)$, respectively. This captures
all possible isometry groups for a general three-sphere, the
latter, most symmetric case corresponding to the usual round
sphere. The Ricci flow describing the renormalization of the
sigma-model leads unavoidably to the round sphere, which is
therefore the unique perturbative infra-red fixed point, found at
one loop.



\section{Space--time solutions}
\label{sec:equations-motion}

\subsection{Equations of motion for the dilaton and the scale factor}

As advertised in the introduction, our aim is now to describe a
generalization of the previous construction obtained by
introducing an extra time dimension and treating the coupling as a
time-dependent field. This clearly bears many resemblances with
the usual Liouville dressing of
\cite{David:1988hj,Polyakov:1998ju,Ellis:1998dw,Ellis:2000dy}. In
other words, we would like to write down the Weyl-invariance
equations for the following sigma model:
\begin{equation}
\label{eq:time-action}
  S = \frac{1}{2} \int \di^2 z \: \left[ - \partial t \bar \partial t
  + \left( c(t) g_{\mu \nu} + B_{\mu \nu}  \right) \partial X^\mu \bar \partial X^\nu + R \Phi (t) \right] ,
\end{equation}
where, $g$ and $B$ are background fields solving the
$d$-dimensional equations of motion.  Modulo the time-dependent
dilaton, the system in Eq.~\eqref{eq:low-dimensional-sigma}
appears thus as a description of a constant-time slice.

Let us start by rewriting the $d+1$ dimensional metric in the form of
a Weyl rescaling,
\begin{equation}
\label{eq:g-bar}
  \bar g_{\textsc{mn}} = \mathrm{e}^{2 \sigma(t)}
  \begin{pmatrix}
    - \mathrm{e}^{-2\sigma(t)} & 0 \\
    0 & g_{\mu \nu}
  \end{pmatrix} = \mathrm{e}^{2 \sigma (t)} g_{\ssc{mn}} ,
\end{equation}
where $c(t) = \mathrm{e}^{2 \sigma (t)}$, as before. This means in
particular that the Ricci tensor can be written as
\begin{equation}
  \overline{R}_{\ssc{mn}} = R_{\ssc{mn}} - g_{\ssc{mn}} K\du{\ssc{l}}{\ssc{l}} -
  \left(d - 1 \right) K_{\ssc{mn}} ,
\end{equation}
where $K_{\ssc{mn}}$ is defined as
\begin{align}
  K\du{\ssc{m}}{\ssc{n}} &= - \partial_{\ssc{m}} \sigma g^{\ssc{nl}} \partial_{\ssc{l}} \sigma
  + g^{\ssc{nl}} \left( \partial_{\ssc{m}} \partial_{\ssc{l}} \sigma -
    \Gamma\ud{\ssc{p}}{\ssc{ml}} \partial_{\ssc{p}} \sigma \right) + \frac{1}{2}
  g^{\ssc{lp}} \partial_{\ssc{l}} \sigma \partial_{\ssc{p}} \sigma
  \delta_{\ssc{m}}^{\phantom{\ssc{m}}\ssc{n}}, \\
  K_{\ssc{mn}} &= g_{\ssc{nl}} K\du{\ssc{m}}{\ssc{l}} .
\end{align}
After some algebra one finds
\begin{subequations}
\label{eq:Ric-bar}
  \begin{align}
    \overline{R}_{tt} &= -d \left( \ddot \sigma (t)+ \dot \sigma^2(t) \right) ,\\
    \overline{R}_{t\mu } &= 0 ,\\
    \overline{R}_{\mu \nu } &= R_{\mu \nu } + \bar g_{\mu \nu} \left( d \dot
      \sigma^2(t) + \ddot \sigma(t) \right) .
  \end{align}
\end{subequations}
The $(d+1)$-dimensional metric is not Einstein, so we expect the
presence of a time-dependent dilaton. The equations of motion
read:
\begin{equation}
  \begin{cases}
    \beta_\Phi = -\frac{1}{2} \bar{\square} \Phi + \bar \nabla_{\ssc{m}}
    \Phi \bar \nabla^{\ssc{m}} \Phi - \frac{1}{24} H_{\ssc{mnp}} \bar H^{\ssc{mnp}}
    = \frac{d+c_\mathrm{I}-25}{6}  ,
    \\
    \beta_G = \overline{R}_{\ssc{mn}} - \frac{1}{4} \bar
    H_{\ssc{mn}}^2 + 2
    \bar \nabla_{\ssc{m}} \bar \nabla_{\ssc{n}} \Phi = 0 ,\\
    \beta_B = - \frac{1}{2} \bar \nabla^{\ssc{m}} H_{\ssc{mnp}}
    + \partial^{\ssc{m}} \Phi H_{\ssc{mnp}} = 0 .
  \end{cases}
\end{equation}

Several remarks are in order here.
\begin{itemize}
  \item As already pointed out, the action in
Eq.~\eqref{eq:time-action} does not describe by itself a critical
string theory. An internal conformal field theory of central
charge $c_{\mathrm{I}}$ must be superimposed to the above
sigma-model. Requiring that we are in the perturbative regime with
respect to $\alpha'$, we expect: (\romannumeral1) the model in
Eq.~\eqref{eq:time-action} to have a central charge of the order
of $d+1$ and (\romannumeral2) the internal \textsc{cft} with
$c_{\mathrm{I}} \approx 25-d$ to have a geometrical interpretation
with a $(25-d)$-dimensional compact target space (we refer here to
the bosonic string).
  \item If the scale factor were chosen to be kept constant and equal
to its critical value $c=1$, the natural solution to the above
equations would be the \emph{linear dilaton} $\Phi = \pm q t$.
This solution is actually an exact \textsc{cft} with central
charge $1-3q^2<1$ (the minus sign is due to the time-like
direction). The central charge $c_d$ of the $d$-dimensional sigma
model (decoupled in this case) depends on the geometry (the
curvature) of the target space. For some spherical-like geometry,
one expects $c_d\lesssim d$ and therefore the internal
\textsc{cft} must have $c_{\mathrm{I}} \gtrsim 25-d$. This imposes
a locally hyperbolic internal target space that can be
\textit{e.g.} a quotient of a hyperbolic plane $H_{25-d}$ by some
discrete subgroup.
\end{itemize}


The $\beta_\Phi = \text{const.}$ equation can be seen as a consequence
of the Bianchi identities and the $\beta_B = 0$ and $\beta_G = 0$
equations. Explicitly it reads:
\begin{equation}
  \beta_\Phi = - \frac{1}{2} \ddot \Phi +
  \dot \Phi \left( \dot \Phi - \frac{d}{2} \dot \sigma \right) -
  \frac{1}{6} \mathrm{e}^{-6\sigma} R = \frac{d+c_\mathrm{I}-25}{6}
\end{equation}
The $\beta_B = 0$ condition is satisfied by our choice of Kalb--Ramond
field (see Sec. \ref{sec:overall-scale})
and a time-dependent dilaton $\Phi = \Phi(t)$. So, let us consider
the remaining $\beta_G=0$ equation. First of all we separate the
time component (all terms retain a block-matrix structure):
\begin{equation}
  \begin{cases}
    R_{tt} + 2 \partial_t \partial_t \Phi(t) = -d \left( \ddot \sigma(t) + \dot \sigma^2(t)
    \right) + 2 \ddot \Phi(t) = 0 ,\\
    R_{\mu \nu} \left( 1 - \mathrm{e}^{-4\sigma(t)} \right) + \bar g_{\mu \nu} \left( d
      \dot \sigma^2(t) + \ddot \sigma(t) - 2 \dot \sigma(t) \dot \Phi(t) \right) = 0
      ,
  \end{cases}
\end{equation}
where we used the equations of motion for the $\sigma=0$ conformal system
\begin{equation}
  \label{eq:space-equations}
  R_{\mu \nu} = \frac{1}{4} H_{\mu\nu}^2 .
\end{equation}
Taking the trace with $\bar g^{\ssc{mn}}$ one obtains the system:
\begin{equation}
  \begin{cases}
    d \left( \ddot \sigma(t)  + \dot \sigma^2(t) \right) - 2 \ddot \Phi(t) = 0 , \\
    R \mathrm{e}^{-2\sigma(t)} \left( 1 - \mathrm{e}^{-4\sigma(t)} \right) + d \left( d \dot
      \sigma^2(t) + \ddot \sigma(t) - 2 \dot \sigma(t) \dot \Phi(t) \right) = 0 .
  \end{cases}
\end{equation}
Introducing
\begin{equation}
  \label{eq:defin-Psi}
  \Psi (t) = -\dot \Phi (t) + \frac{d}{2} \dot \sigma (t) ,
\end{equation}
the equations become:
\begin{equation}
\label{eq:friction-syst}
  \begin{cases}
    \dot \Psi(t) = - \frac{d}{2} \dot \sigma^2(t) ,\\
    \ddot \sigma(t) = - \frac{R}{d} \mathrm{e}^{-2\sigma(t)} \left( 1 - \mathrm{e}^{-4 \sigma(t)} \right) - 2
    \dot \sigma(t) \Psi(t) .
  \end{cases}
\end{equation}

The second equation has precisely the structure of the equation of
motion of a point particle in a potential
\begin{equation}\label{eq:pot}
  V (\sigma) = \frac{R}{6d} \mathrm{e}^{-6 \sigma} \left( 1 - 3 \mathrm{e}^{4\sigma} \right) + \mathrm{cst.} ,
\end{equation}
in the presence of a time-dependent friction coefficient
$\Psi(t)$. In the $\Psi \to \infty $ limit we recover the same
equation as in Eq.~\eqref{eq:energy-evol-sigma} with the same
potential $V(\sigma)$ when identifying the \emph{energy scale}
$\mu$ for the running, off-shell (\textit{i.e.} non-conformal),
$d$-dimensional system with the \emph{time direction} according to
\begin{equation}
  \log \mu = - \frac{2\pi  t}{\bar \Psi} .
\end{equation}
This was to be expected since what we find is the usual link between
the \textsc{rg}-flow equations and the conformal condition typical of
a Liouville dressing (see \emph{e.g}
\cite{David:1988hj,Distler:1988jt,Polyakov:1998ju,Ellis:2000dy}).
This identification holds, however, only in the large-$Q$ regime.

A useful reformulation of the equations is obtained by multiplying the
second equation by $\dot \sigma$ and introducing the energy $E = \dot
\sigma^2 / 2 $:
\begin{equation}
  \begin{cases}
    \dot \Psi(t) = - d E(t) ,\\
    E(t) = - V(\sigma(t)) + \frac{2}{d} \Psi^2 (t) ,
  \end{cases}
\end{equation}
which in turn allows us to write down an equation for $\Psi $ as a
function of $\sigma$:
\begin{equation}
  \frac{\di}{\di \sigma} \Psi (\sigma) = -d \sqrt{ \frac{\Psi(\sigma)^2}{d} - \frac{V(\sigma)}{2}} .
\end{equation}

\subsection{Linearization}

\begin{figure}
  \begin{center}
    \subfigure[Small $\Psi(0)$]{
      \includegraphics[width=.45\linewidth]{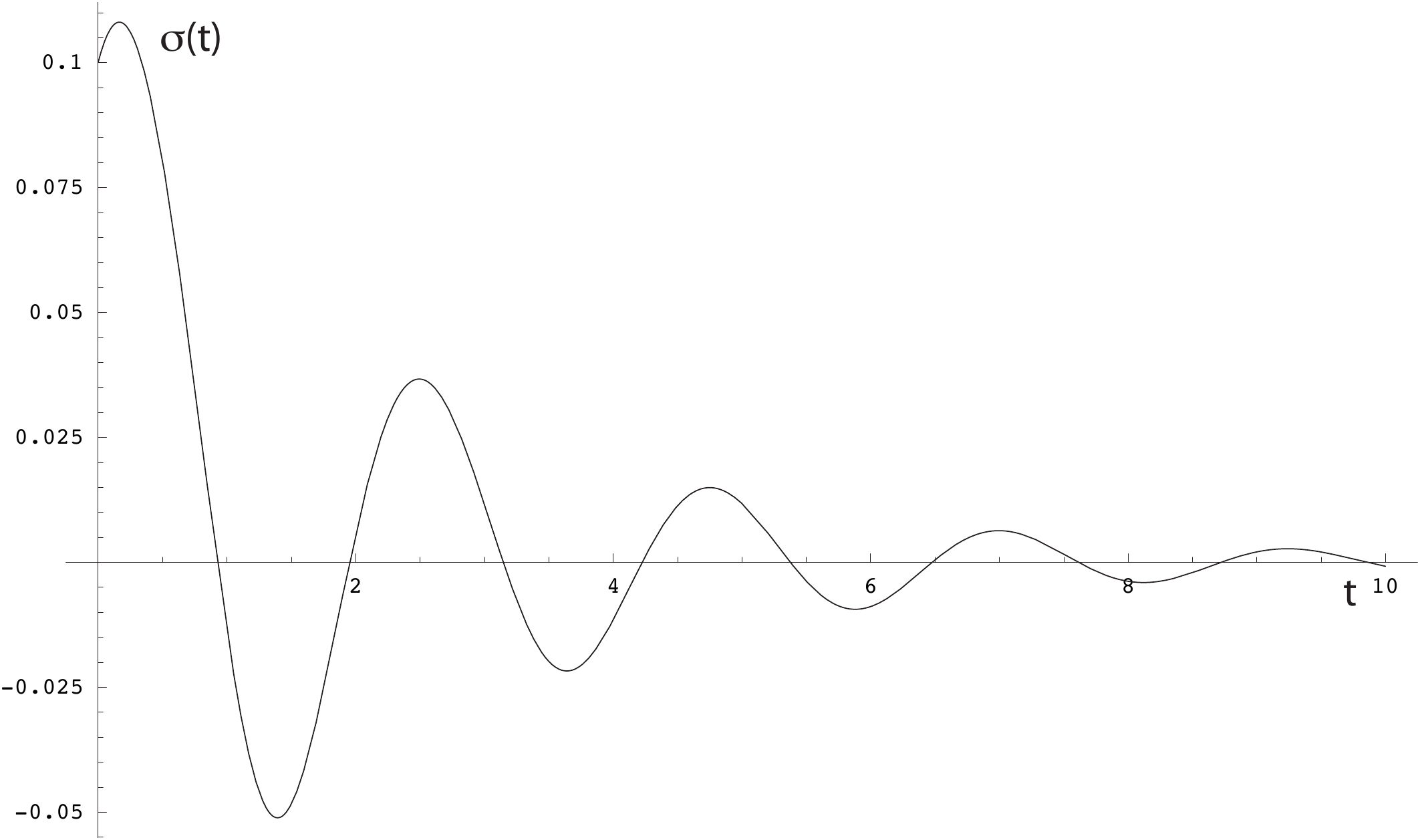}}
    \subfigure[Large $\Psi(0)$]{
      \includegraphics[width=.45\linewidth]{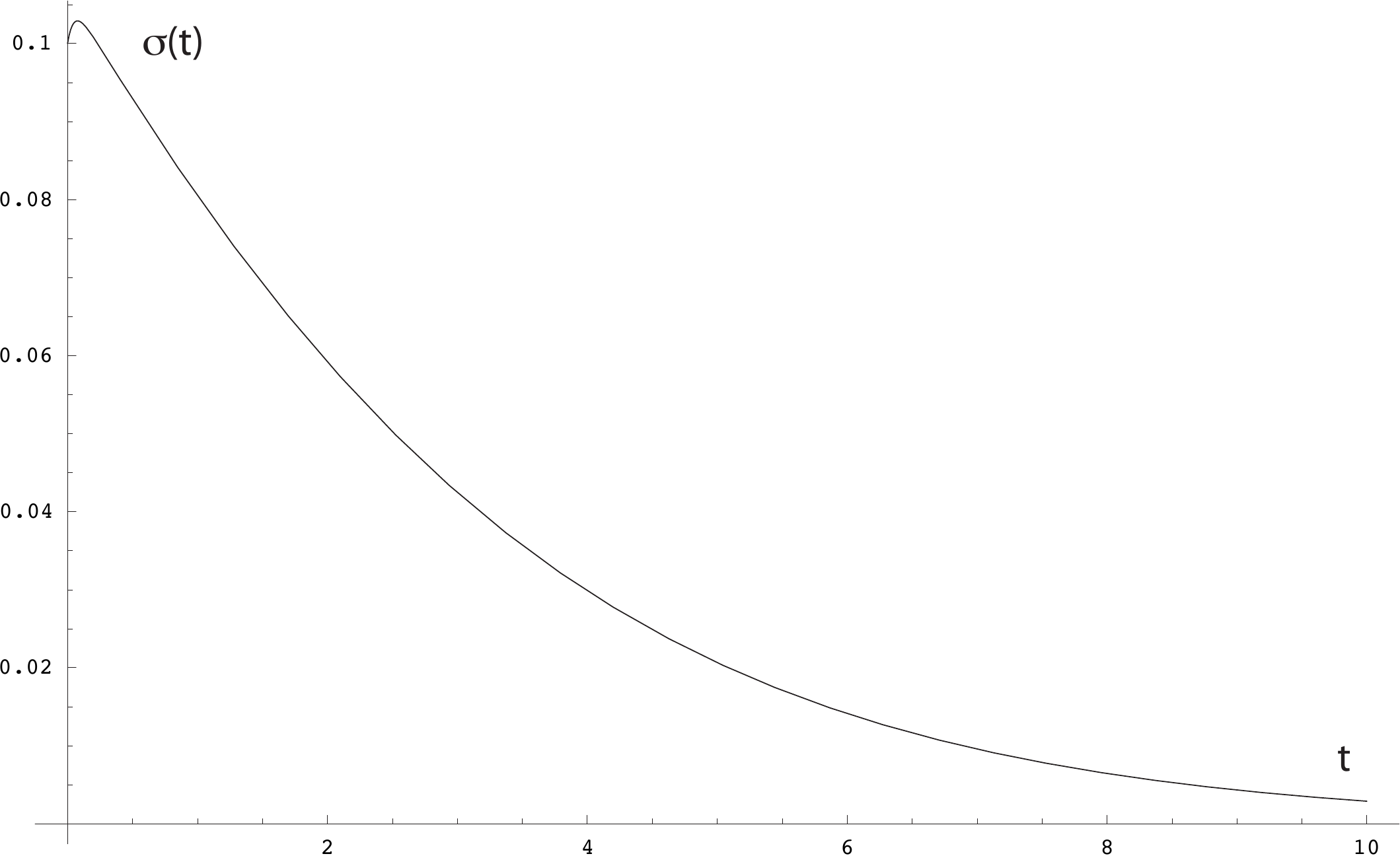}}
  \end{center}
  \caption{Typical behaviour for $\sigma(t)$ in the non-linear system. For (a)~small and (b)~large
    (positive) initial values of $\Psi(t)$ (numerical integration with
   (a) $\left( \Psi (0), \sigma(0), \dot \sigma(0) \right) = \left(
    0.5, 0.1, 0.1) \right)$ and (b) $\left( \Psi (0), \sigma(0), \dot \sigma(0)    \right) =
    \left( 10, 0.1, 0.1\right)$).}
  \label{fig:num-solution-Psi}
\end{figure}

The system~\eqref{eq:friction-syst} can be solved numerically and
typical results for large $\Psi (0)$ and small $\Psi(0)$ are shown in
Fig.~\ref{fig:num-solution-Psi}. Since we describe a dissipative
motion, the system relaxes to equilibrium with or without
fluctuations.

The solutions we found are in principle only valid at first order
in $\alpha'$ so one may wonder about their stability. Actually, it
turns out that if the spatial part is supersymmetric (which is the
case if the model in Eq.~\eqref{eq:low-dimensional-sigma} is of
the Wess-Zumino-Witten type), then so is the spacetime in
Eq.~\eqref{eq:g-bar}. This is immediate once one notices that the
Weyl tensor for $\bar g$ is the same as for the Cartesian product
with line element $\di s^2 = -\di t^2 + g_{\mu \nu} \di x^\mu \di
x^\nu$ and therefore vanishes identically.

\paragraph{Around the conformal point.}

A further step can be made via linearization. To study the behaviour
in a neighbourhood of the $\sigma = 0$ conformal point we introduce
\begin{equation}
  \Sigma (t)= \dot \sigma (t)
\end{equation}
and derive the first-order system
\begin{equation}
  \begin{cases}
    \dot  \Psi(t) = - \frac{d}{2} \Sigma^2(t) ,\\
    \dot \sigma(t) = \Sigma(t) ,\\
    \dot \Sigma(t) = - V^\prime (\sigma(t)) - 2 \Sigma(t) \Psi(t) ,
  \end{cases}
\end{equation}
which has a fixed point for $\left( \Psi(t), \sigma(t), \Sigma(t)
\right) = \left( \bar \Psi, 0, 0 \right)$, $\bar \Psi$ being a
constant. Around this point the asymptotic behaviour is described by
\begin{equation}
  \label{eq:linear-system}
  \begin{cases}
    \dot \Psi(t) = 0 ,\\
    \dot \sigma(t) = \Sigma(t) ,\\
    \dot \Sigma(t) = - V^{\prime \prime} (0) \sigma(t) - 2 \bar \Psi \Sigma(t) = - \frac{4 R}{d}
    \sigma(t) - 2 \bar \Psi \Sigma(t) .
  \end{cases}
\end{equation}
$\Psi $ decouples (and remains constant) and the only non-trivial
remaining equation of motion is
\begin{equation}
  \frac{\di^2 \sigma (t)}{\di t^2} = - \frac{4 R}{d}
  \sigma(t) - 2 \bar \Psi \frac{\di \sigma (t)}{\di t} ,
\end{equation}
which can be integrated to give
\begin{subequations}
\label{eq:linear-solution}
  \begin{gather}
    \sigma (t) = C_1 \exp \left[ -\left( \bar \Psi + \sqrt{\bar \Psi^2
          - \frac{4R}{d}} \right) t \right] + C_2 \exp \left[ -\left(
        \bar \Psi - \sqrt{\bar \Psi^2 - \frac{4R}{d}} \right)
      t \right] ,\\
    \Phi (t) = \Phi_0 + \frac{dC_1}{2} \exp \left[ -\left( \bar \Psi +
        \sqrt{\bar \Psi^2 - \frac{4R}{d}} \right) t \right] +
    \frac{dC_2}{2} \exp \left[ -\left( \bar \Psi - \sqrt{\bar \Psi^2 -
          \frac{4R}{d}} \right) t \right] - \bar \Psi t,
  \end{gather}
\end{subequations}
where $C_1 $ and $C_2$ are integration constants.

For positive $\bar \Psi$ the solution converges to $\sigma = 0$ with
or without oscillations depending on whether $\bar \Psi^2 \lessgtr 4R/d$, which
clarifies the meaning of ``large'' and ``small'' $\Psi(0)$ in
Fig.~\ref{fig:num-solution-Psi}.

In terms of $\sigma(t) $ and $\Phi(t)$, this limit solution is
\begin{align}
  \sigma(t) \xrightarrow[t\to \infty ]{} 0 && \Phi (t) \sim \Phi_0 - \bar \Psi t ,
\end{align}
which is the conformal model in Eq.~\eqref{eq:low-dimensional-sigma}
plus a linear dilaton.

\paragraph{Large $\sigma$.}

In the large-$\sigma$ limit, the solution goes to flat space. To
study its behaviour let us consider the system
\begin{equation}
  \begin{cases}
    \dot \Psi (t) = - d \left( - V(\sigma) + \frac{2}{d} \Psi(t)^2 \right) ,\\
    \ddot \sigma(t) = - V^\prime(\sigma(t)) - 2 \dot \sigma(t) \Psi(t) .
  \end{cases}
\end{equation}
In the large-$\sigma$ limit $V(\sigma) \to C $ and $V^\prime (\sigma)
\to 0$ so it reduces to
\begin{equation}
  \begin{cases}
    \dot \Psi (t) = - d \left( -C + \frac{2}{d} \Psi(t)^2 \right) ,\\
    \ddot \sigma(t) = - 2 \dot \sigma(t) \Psi(t) ,
  \end{cases}
\end{equation}
which can be solved analytically:
\begin{align}
  \Psi (t) &= \sqrt{\frac{Cd}{2}} \tanh \left(2 \sqrt{C d}  t + C_1 \right) ,\\
  \sigma (t) &= C_3 + C_2 \arctan \left( \sinh C_1 + \cosh C_1 \tanh \left(\sqrt{\frac{C d}{2}} t\right)\right) \\
  \Phi (t) &= \Phi_0 - \frac{1}{2\sqrt{2}} \log \left( \cosh \left(2 \sqrt{C d}  t + C_1 \right) \right) + \frac{d}{2 } \sigma(t)
\end{align}
where $C_1,C_2,C_3$ are integration constants. As expected, for large
$t$, both $\Psi (t)$ and $\sigma(t)$ asymptotically converge to
constant values and the background is flat with a linear dilaton.

\subsection{The meaning of $\bar \Psi$}

The parameter $\bar \Psi$ is interpreted as the background charge
that drives the dilaton, in the regime where the latter is linear
with time. This parameter dictates the behaviour of the solution
around the fixed point in many ways. Apart from discriminating
between exponentially decreasing and oscillating solutions, it
also measures the consistency of the perturbative string approach.
In fact positive values of $\bar \Psi$ correspond to a dilaton
evolving towards large negative values, \emph{i.e.} towards the
perturbative regime. On the other hand, negative values of $\bar
\Psi$ would lead to diverging solutions in which we would lose
control over the approximations we made. It is worth to remark
that in the non-linear dynamics $\Psi (t)$ is not allowed to
change its sign if we make the hypothesis of unicity of the
solution. In fact changing sign would require to cross the $\Psi =
0, \sigma = \text{const.}$ line corresponding to the Euclidean
conformal model.

A final remark regards the consistency of the approximation for
the dynamics one obtains from the \textsc{rg}-flow
equation~\eqref{eq:energy-evol-sigma}, corresponding to a $\Psi
\to \infty$ limit. The linearized system~\eqref{eq:linear-system}
provides a justification for such limit: in fact the time scale
for $\Psi (t)$ is comparably larger than the time scale for
$\sigma(t)$ -- to the point that the former decouples around the
fixed point. For this reason it can be taken as a constant (fixed
by the initial conditions) if we just concentrate on the evolution
of the warp factor $\sigma (t)$.

\subsection{The anisotropic deformations of the three-sphere}

We will now study the anisotropic deformation of the $SU(2)$ model
(see Sec. \ref{sec:three-sphere}) when an extra time dimension is
added to the system while the parameters $\gamma_\nu$ (Eq.
(\ref{eq:defmets})) and the dilaton $\Phi$ are allowed to depend
on it. Explicitly, the full four-dimensional, time-dependent
metric under consideration is
\begin{multline}
  \di s^2 = - \di t^2 + \sum_{\nu=1}^3 \gamma_\nu (t) J^\nu J^\nu =
  -\di t^2 + k\Big( \left(
    \gamma_2 (t) \cos^2 \phi + \gamma_1 (t) \sin^2 \phi \right) \di
  \theta^2 + \\
  +  \left( \gamma_1(t) \cos^2 \theta \cos^2 \phi + \gamma_3(t) \sin^2
    \theta + \gamma_2(t) \cos^2 \theta \sin^2 \phi \right) \di
  \psi^2 + \\
  + \gamma_3(t) \di \phi^2 + 2 \gamma_3 (t) \sin \theta \, \di \phi \, \di
  \psi + 2 \left( \gamma_1(t) - \gamma_2 (t) \right) \cos \theta \cos
  \phi \sin \phi \, \di \theta \, \di \psi\Big).
\end{multline}
 We introduce again the functions $\sigma_\nu(t)$ as
\begin{equation}
  \gamma_\nu (t) = \mathrm{e}^{2 \sigma_\nu(t)},
\end{equation}
and find the following equations of motion:
\begin{equation}
  \begin{cases}
    2 \ddot \Phi (t) =\displaystyle{ \sum_{\nu=1}^3 \left( \ddot
        \sigma_\nu(t) + \sigma_\nu(t)^2 \right)}, \\
    \displaystyle{\ddot \sigma_1(t) =  \frac{R}{3} \left(1 - \mathrm{e}^{4 \sigma_1(t)} + \left( \mathrm{e}^{2
            \sigma_2(t)} - \mathrm{e}^{2 \sigma_3(t)}
        \right)^2 \right)\prod_{\nu=1}^3 \mathrm{e}^{-2
        \sigma_\nu(t)}- \dot \sigma_1(t) \left(  - 2\dot \Phi(t) + \sum_{\nu=1}^3
        \sigma_\nu(t)\right)},
  \end{cases}
\end{equation}
plus permutations, where $R=\nicefrac{6}{k}$ is the curvature of
the undeformed  three-sphere. Here, it is natural to define
\begin{equation}
  Q (t) = - \dot \Phi (t) + \frac{1}{2} \sum_{\nu=1}^3 \dot \sigma_\nu (t)
\end{equation}
and recast the system in the form
\begin{equation}
  \begin{cases}
    \displaystyle{\dot Q(t) = - \frac{1}{2} \sum_{\nu=1}^3 \dot \sigma_\nu^2 (t)}, \\
    \displaystyle{\ddot \sigma_1(t) = \frac{R}{3}
    \left(  1- \mathrm{e}^{4 \sigma_1(t)} + \left( \mathrm{e}^{2 \sigma_2(t)}
   - \mathrm{e}^{2
          \sigma_3(t)} \right)^2 \right)
           \prod_{\nu=1}^3 \mathrm{e}^{-2 \sigma_\nu(t)}- 2 \dot \sigma_1(t)
          Q(t)},
  \end{cases} \label{eq:anisev}
\end{equation}
plus permutations.

The analysis of  Eqs. (\ref{eq:anisev}) can be performed in
analogy with the system we obtained above, when considering only
the overall breathing mode. We first observe that given the
\textsc{rg}-flow equations\footnote{Notice that $\beta_\nu
(\mathbf{\sigma})=\nicefrac{\beta_\nu
(\mathbf{\gamma})}{2\gamma_\nu}$.}
\begin{equation}
  \frac{\di \sigma_\nu (\mu)}{\di \log \mu} = \beta_\nu (\mathbf{\sigma}
  (\mu)),
\end{equation}
we obtain those for the time-dependent system as
\begin{equation}
  \frac{\di^2 \sigma_\nu(t)}{\di t^2} = \beta_\nu (\mathbf{\sigma}(t))
  - 2 Q(t) \frac{\di \sigma_\nu(t)}{\di t}
\end{equation}
where $Q(t)$ is the charge that naturally generalizes the one we
found before. From these we conclude that the structure is again
that of a dissipative system with the time-dependent friction
coefficient given by the charge. We can further proceed by
linearization around the $SU(2)$ \textsc{wzw} fixed point. This
goes along the same lines as in the isotropic case and gives
\begin{equation}
  \begin{cases}
    \dot Q (t) = 0, \\
    \ddot \sigma_\nu (t) = - \frac{4R}{d} \sigma_\nu (t)  -
    2 \bar Q \dot \sigma_\nu (t), \hspace{2em} \nu = 1, 2,
    3.
  \end{cases}
\end{equation}
The charge $Q(t)$ decouples and remains constant $Q(t) = Q(0) =
\bar Q$ while the three modes become independent from each other
and obey the same differential equation. This means that they
oscillate in phase with each other and eventually converge to the
fixed point corresponding to $SU(2)_k$ \textsc{wzw} plus a linear
dilaton.

All the considerations we made for the breathing mode then extend
directly to this anisotropic generalization. The net effect of the
time dependence is therefore to wash out the anisotropy, as was
expected from the pure \textsc{rg}-flow analysis of Sec.
\ref{sec:three-sphere}.



\section{Cosmological interpretation}
\label{sec:friedm-roberts-walk}

The type of backgrounds we are studying are time-dependent and as such
can be of cosmological interest. For this reason, since there is a
non-trivial dilaton, one should move to the Einstein frame (as opposed
to the string frame we have been using thus far). The metric takes the
form
\begin{equation}
  \tilde g_{\ssc{mn}} = \mathrm{e}^{- \Phi (t) /2 } \bar g_{\ssc{mn}} ,
\end{equation}
and after a coordinate change
\begin{equation}
  \tau(t) = \int_0^t  \mathrm{e}^{-\Phi(t^\prime)/4} \di t^\prime ,
\end{equation}
it reduces to the same warped product form as in
Eq.~\eqref{eq:g-bar}:
\begin{equation}
  \label{eq:Einst-metric}
  \widetilde{\di s^2} = \tilde g_{\ssc{mn}} \di x^{\ssc{m}} \di x^{\ssc{n}} = -\di \tau^2 + \left.
  \mathrm{e}^{2 \sigma(t) - \Phi (t)/2}
  \right|_{t = t(\tau)}  g_{\mu\nu} \di x^\mu \di x^\nu  = -\di \tau^2 + a(\tau)
   g_{\mu\nu} \di x^\mu \di x^\nu
\end{equation}
which has Hubble parameter:
\begin{equation}
  H( \tau) = \frac{\di}{\di \tau} \log (a(\tau))  =  \left. \mathrm{e}^{ \Phi(t)/4} \left( 2 \dot \sigma(t) -
      \frac{\dot \Phi (t)}{2} \right)\right|_{t = t (\tau)} .
\end{equation}

Cosmologically interesting solutions are obtained when $d=3$. In
this case the Kalb--Ramond field $H_{[3]}$ is proportional to the
volume form on $g$. This implies that $H_{\mu \nu}^2 \propto
g_{\mu \nu}$ and then Eq.~\eqref{eq:space-equations} reduces to
\begin{equation}
  R_{\mu \nu} = 2\Lambda g_{\mu \nu} ,
\end{equation}
\emph{i.e.} $g_{\mu \nu}$ is the metric of an Einstein
three-manifold. The simplest case is a three-sphere for which we
get a typical example of \textsc{frw} space--time of the kind
studied in
\cite{Tseytlin:1991ss,Tseytlin:1992ye,Goldwirth:1993ha,Copeland:1994vi}.
As such it describes the time evolution of a homogeneous and
isotropic\footnote{We will discuss in this section the isotropic
situation only.} space--time, or more generally of a space--time
with the target-space symmetries of the conformal theory in
Eq.~\eqref{eq:low-dimensional-sigma}.

Some intuition about the time evolution can be developed if we
take the linearized system in Eq.~\eqref{eq:linear-system}.
\paragraph{Linear Dilaton.} In fact, as remarked above, if we
consider the large $t$ limit, the solution asymptotically
approaches a linear dilaton background (which was already studied
from this point of view in \cite{Antoniadis:1988vi}):
\begin{align}
  \sigma (t) \xrightarrow[t\to \infty ]{} 0 && \Psi (t) = \bar \Psi && \Phi(t) \sim - \bar \Psi t ,
\end{align}
hence one can easily verify that the metric in the Einstein frame is
asymptotically
\begin{equation}
  \widetilde{\di s^2} \sim - \di \tau^2 + \bar \Psi^2 \tau^2g_{\mu\nu} \di x^\mu \di x^\nu ,
\end{equation}
which corresponds to an expanding universe with Hubble parameter
\begin{equation}
  H( \tau) = \frac{2}{\tau}
\end{equation}
and curvature
\begin{equation}
  \tilde R \sim \frac{R + \bar \Psi^2 d \left( d - 1 \right)}{\bar \Psi^2 \tau^2} . 
\end{equation}

\paragraph{Exponentially decreasing $\sigma$.}

A similar result with a polynomial expansion is found if we consider
an exponential decrease for $\sigma(t)$, or better for $c(t)$ (in the
linear limit $c(t) - 1$ obeys the same equations as $\sigma(t)$)
\emph{i.e.} if we are in the large-$\bar \Psi$ regime. After a
redefinition of the variables we can write:
\begin{equation}
  c(t) = \mathrm{e}^{-\alpha t} + 1  .
\end{equation}
It is easy to check that in general\footnote{Since $c(t)>0$ by
  construction, the relation $\tau = \tau (t)$ is always invertible.}
\begin{equation}
  \tau (t) = \int c(t)^{-d/16} \mathrm{e}^{1/4\int \Psi(t^\prime) \di t^\prime} \di t ,
\end{equation}
and in this linearized approximation this becomes
\begin{equation}
  \tau (t)= \int \left( \mathrm{e}^{- t} + 1 \right)^{-d/16} \mathrm{e}^{\bar \Psi t / 4} \di t .
\end{equation}
Although this integral can be solved analytically:
\begin{equation}
  \tau (u) = \frac{16}{d + 4 \bar \Psi} \left( 1 + \frac{1}{u} \right)^{-d/16} u^{\bar \Psi/4} \left( 1 + u \right)^{d/16} {}_2F_1 \left( \frac{d}{16} , \frac{d+4 \bar \Psi}{16}; \frac{d + 4 \bar \Psi}{16} + 1, - u\right) ,
\end{equation}
where $u = \mathrm{e}^t$ and ${}_2F_1$ is an hypergeometric
function\footnote{The hypergeometric function ${}_2F_1$ is defined
as
  follows:
  \begin{equation}
    {}_2F_1 (a,b;c,z) =
    \sum_{k=0}^\infty \frac{ \left( a \right)_k \left( b \right)_k}{\left(c \right)_k}
    \frac{z^k}{k!},
  \end{equation}
  where $\left(a \right)_k$ is the Pochhammer symbol
  \begin{equation}
    \left( a \right)_k = \frac{\Gamma (a+k)}{\Gamma (a)}.
  \end{equation}%
}. It is better to consider the asymptotic behaviour. For $u \to
\infty$ one finds that $\tau (u)$ and the warp factor $a(u)$ go
as:
\begin{align}
  \tau (u) \underset{\tau \to \infty}{\sim} \frac{4}{\bar \Psi} u^{\bar \Psi /4} ,
  && a(u) \underset{\tau \to \infty}{\sim} u^{\bar \Psi /2} ,
\end{align}
and consistently with the results above for the linear dilaton case (which is
precisely the large-$u$ limit):
\begin{equation}
  a(\tau) \underset{\tau \to \infty}{\sim} \tau^2 ;
\end{equation}
similarly for small $u$:
\begin{align}
  \tau (u) \underset{\tau \to 0}{\sim} \frac{16}
  {d+4 \bar \Psi} u^{\left( d+4 \bar \Psi\right)/16} , &&
  a(u) \underset{\tau \to 0}{\sim} u^{d/4+2+\bar \Psi/2}
\end{align}
and then
\begin{equation}
  a(\tau) \underset{\tau \to 0 }{\sim}
  \tau^{4 + 8 \left( 4 - \bar \Psi \right)/\left( d + 4 \bar \Psi \right)} .
\end{equation}
Since we have a power-law expansion at both limits, the Hubble
parameter goes as $H(\tau) \propto 1/ \tau$ and more precisely is
asymptotically given by:
\begin{align}
  H(\tau) \underset{\tau \to \infty}{\sim} \frac{2}{\tau} && H(\tau) \underset{\tau \to 0}{\sim} \frac{2}{\tau} \left(1 + \frac{ 16 + d}{ d + 4 \bar \Psi }  \right) .
\end{align}

Note that this behaviour precisely measures the effect of a finite
value for $\bar \Psi$ and in fact for $\bar \Psi \to \infty$ we
recover again $H (\tau) \sim 2/\tau$.  To summarize, we get again a
polynomially expanding universe (a so-called big-bang solution).

\paragraph{Damped oscillation regime.}

The small-$\bar \Psi$ regime is more difficult to be studied
analytically. Nevertheless we can make some qualitative comments on the overall behaviour of the solution.

\begin{figure}
  \begin{center}
    \subfigure[Warp factor $a(\tau)$]{
      \includegraphics[width=.45\linewidth]{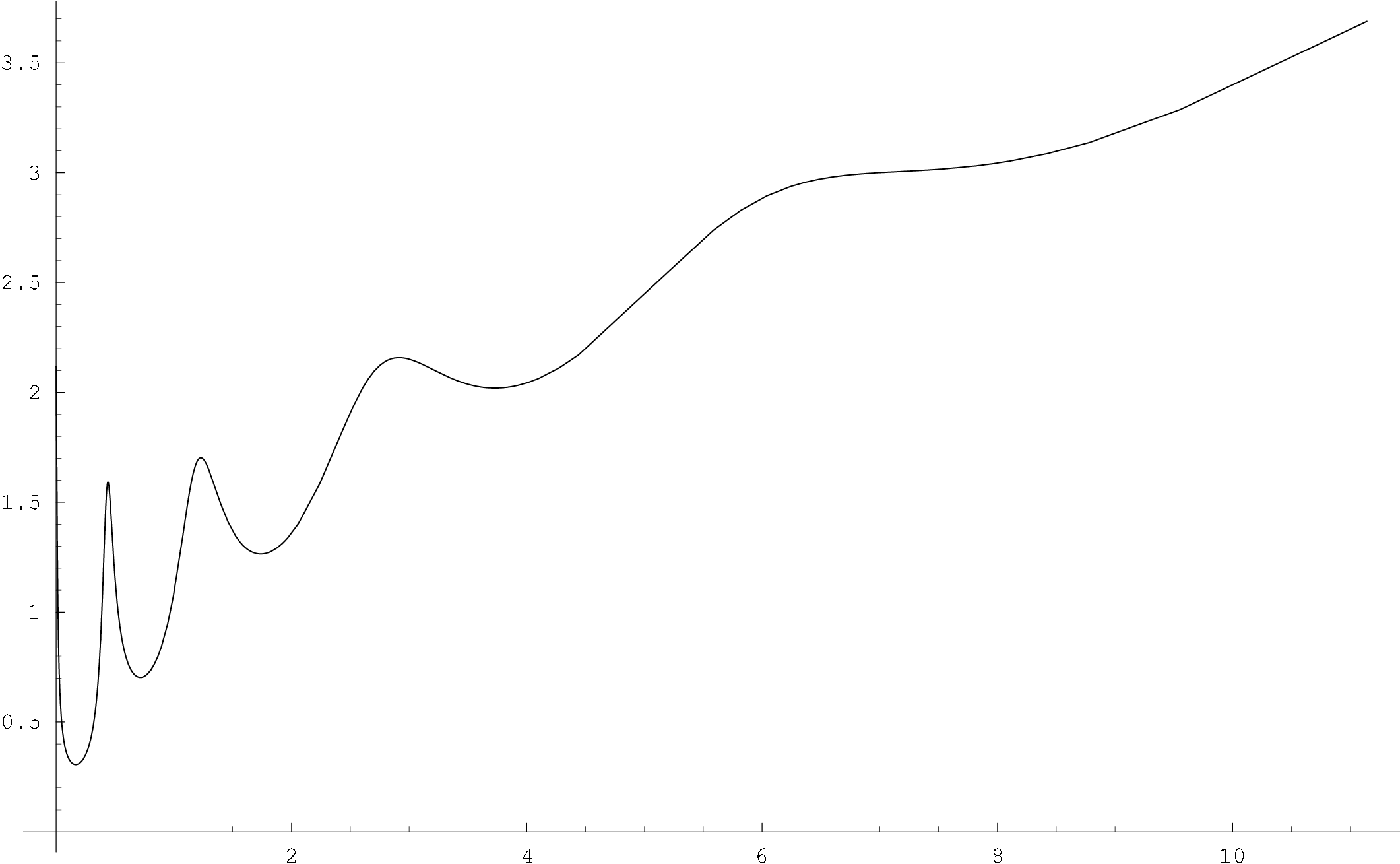}}
    \subfigure[Hubble parameter $H(\tau)$]{
      \includegraphics[width=.45\linewidth]{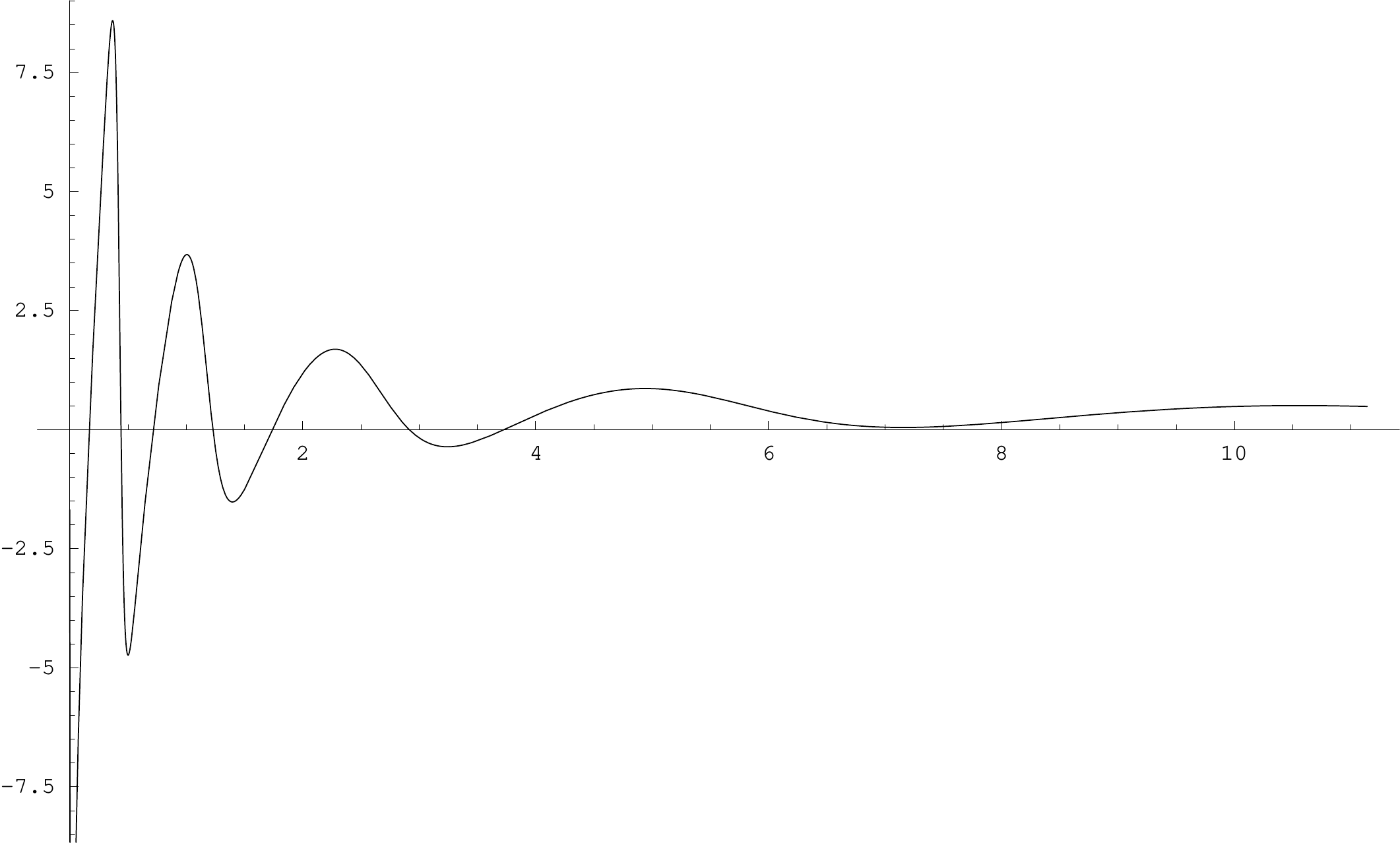}}
  \end{center}
  \caption{Typical behaviour for the warping factor and Hubble
    parameter in the small-$\bar \Psi$ regime. Numerical solution with parameters $A=1, \bar \Psi = 1.2, \omega = 10, d =3$. }
  \label{fig:num-solution-warp}
\end{figure}

First of all we can redefine the variables so to recast Eqs.~\eqref{eq:linear-solution} in the form
\begin{subequations}
  \begin{gather}
    \sigma (t) = A \mathrm{e}^{- \bar \Psi t} \cos ( \omega t ) \\
    \Phi (t) = \Phi_0 + \frac{A d}{2} \mathrm{e}^{-\bar \Psi t} \cos (\omega t) - \bar \Psi t
  \end{gather}
\end{subequations}
where $A$ is a constant and $\omega^2 = \abs{\bar Q^2 - 4 R/d}$. The
warp factor and the Hubble parameter then read
\begin{gather}
  a(\tau ) = \left. \mathrm{e}^{2 \sigma(t) - \Phi(t)/2} \right|_{t = t (\tau)} = \left.
  \exp \left[ A \mathrm{e}^{-\bar \Psi t} \cos (\omega t) \left(2 - \frac{d}{4} \right) -
  \frac{\Phi_0}{2} + \frac{\bar Q t}{2} \right] \right|_{t = t (\tau)}, \\
  \begin{split}
    H(\tau ) = \left. \mathrm{e}^{ \Phi(t)/4} \left( 2 \dot \sigma(t) -
        \frac{\dot \Phi (t)}{2} \right)\right|_{t = t (\tau)} =
    \exp \left[ \frac{\Phi_0}{4} + \frac{A d}{8} \mathrm{e}^{-\bar \Psi
          t} \cos (\omega t) - \frac{\bar \Psi t}{4} \right] \times \\ \times  \left. \left( -
        A \mathrm{e}^{-\bar \Psi t} \left( \bar \Psi \cos (\omega t) + \omega
          \sin (\omega t) \right) \left(2 - \frac{d}{4} \right) +
        \frac{\bar \Psi}{2} \right) \right|_{t = t (\tau)},
  \end{split}
\end{gather}
where $\tau (t)$ is defined by
\begin{equation}
  \tau (t) = \int_0^t \mathrm{e}^{-\Phi (t^\prime)/4} \di t^\prime = \int_0^t \exp \left[ - \frac{\Phi_0}{4} - \frac{A d}{8} \mathrm{e}^{-\bar \Psi t^\prime} \cos (\omega t^\prime) + \frac{\bar \Psi t^\prime}{4} \right] \di t^\prime .
\end{equation}
The Einstein frame time variable $\tau (t)$ is given by the integral
of a positive function of $t$. This means that it is monotonically
increasing and the relation $\tau = \tau (t)$ is invertible. In turn
the qualitative behaviour of the functions $a = a(\tau)$ and $H =
H(\tau)$ can be understood by studying their behaviours as functions
of the string frame time $t$. In particular $H(\tau)$ is an
oscillating function and has a zero for each zero of the function
$\left(2 \dot \sigma(t) - \dot \Phi (t) /2 \right)$. The maximum
number $\bar N$ of such zeros can be estimated in the following
way. If we write explicitly the equation $2 \dot \sigma - \dot \Phi /2
= 0$ we obtain
\begin{equation}
  \label{eq:H-zeros}
  \bar \Psi = 2 A \mathrm{e}^{-\bar \Psi t} \left(\bar \Psi \cos (\omega t) + \omega \sin (\omega t) \right) \left(2 - \frac{d}{4} \right)  ,
\end{equation}
so, we are looking for the intersections of the line $\varpi(t) =
\bar \Psi$ with the function $\varpi(t) = 2 A \mathrm{e}^{-\bar
\Psi t} \left(\bar \Psi
  \cos (\omega t) + \omega \sin (\omega t) \right) \left(2 -
  \frac{d}{4} \right)$. The latter is bounded by
\begin{equation}
  \varpi(t) = 2 A \mathrm{e}^{-\bar \Psi t} \sqrt{\bar \Psi^2 + \omega^2} \abs{2 - \frac{d}{4}}
\end{equation}
and then intersections are only possible for $t < \bar t$ where $\bar
t $ is given by
\begin{equation}
  \bar t = \frac{1}{\bar \Psi} \log \frac{A \sqrt{\bar \Psi^2 + \omega^2} \abs{8 -d}}{2 \bar \Psi} .
\end{equation}
In the $\left(0, \bar t \right)$ interval, the $\left( \bar \Psi \cos
  (\omega t) + \omega \sin (\omega t) \right)$ term can oscillate $\omega
\bar t / \left(2 \pi \right)$ times and then the number of solutions
for Eq.~\eqref{eq:H-zeros} is at most
\begin{equation}
  n < \bar N = \left\lceil \frac{\omega \bar t}{\pi} \right\rceil  = \left\lceil \frac{\omega}{\bar \Psi \pi} \log \frac{A \sqrt{\bar \Psi^2 + \omega^2}\abs{8-d}}{2 \bar \Psi}  \right\rceil .
\end{equation}
The overall behaviour for the warp factor $a (\tau)$ is of that of a
power-law expansion superimposed with $n < \bar N$ oscillations. The
Hubble parameter after $n$ oscillations goes to zero for large $\tau $
as $H(\tau) \underset{\tau \to \infty }{\sim} 2/\tau$. Typical
behaviours for $a(\tau)$ and $H(\tau)$ are shown in
Fig.~\ref{fig:num-solution-warp}.

\section{Summary}
\label{sec:conclusions}

Finding viable time-dependent solutions with cosmological interest
is one of the open problems in string theory. In this note we have
presented a family of such solutions, derived as deformations of a
conformal model, leading to a generalization of \textsc{frw}
universe undergoing a big bang-like expansion superposed, in some
regimes, with damped oscillations. In these backgrounds the time
dependence comes from a Liouville-type mechanism in which flowing
from the conformal point is seen as the effect of a time-dependent
field. The \textsc{rg}-flow equations on constant-time slices then
become an approximation for the Weyl-invariance equations and the
off-shell dynamics is interpreted as the on-shell dynamics with
one extra dimension. As already stressed, this behaviour is not
generic and the identification of the \textsc{rg}-flow with time
evolution is not valid in any regime.

The constraints imposed by the rational nature of the underlying
compact model essentially fix the type of deformations one can
consider and naturally lead to an isotropic \textsc{frw}
background whose behaviour is due to the presence of an axion
(dual to the Kalb--Ramond field) and an oscillating dilaton.
Anisotropic extensions of \textsc{frw} solution are also
considered but time evolution smoothes the initial anisotropy and
the system relaxes to a symmetric phase.

A common feature of all situations considered here is that the
\textsc{rg} equations at lowest order in $\alpha'$ always describe
a Ricci flow with torsion. In the case of the three-sphere with
anisotropic perturbation, the Ricci-flow equations are identical
to the Halphen system, and are thereby integrable by using modular
forms. More general perturbations exist, which are compatible with
the mini-superspace approximation (see \cite{Fateev:1996ea}) but a
systematic analysis of all those is beyond the scope of this work.

It stems from the overall analysis performed in the present
framework that in a quite universal manner, sooner or later, time
evolution is identified with an \textsc{rg} flow, which is a Ricci
flow. Time becomes thus a two-dimensional scale, the dilaton being
the interplay between target space and worldsheet. Thurston's
geometrization conjecture is then at work. It tells us that the
string target space will universally converge towards a collection
of Einstein spaces, whenever these are available, enhancing
therefore the symmetry. This behaviour is independent of the
initial conditions.

Clearly, this reasoning is valid under the assumption that the
initial space is locally homogeneous. Exact time-dependent string
backgrounds with non-homogeneous spaces exist
\cite{Petropoulos:2006py,Petropoulos:2006pp}, and there time
evolution does not lead to Einstein spaces. Consistently with the
above picture, in these cases, time evolution never identifies
with an \textsc{rg} flow because it is rather related to a
marginal deformation.

The latter observations are puzzling additions to the already
mysterious origin of time in string theory, along the lines of
thought of A. Polyakov.


\bigskip

\begin{acknowledgments}
  The authors thank C.~Bachas, T.~Damour, G.~Gibbons,
  P.~Grange, C.~Kounnas, J.~Rizos and
  K.~Sfetsos for stimulating scientific discussions. This work was
  supported in part by the EU under the contracts MEXT-CT-2003-509661,
  MRTN-CT-2004-005104, MRTN-CT-2004-503369. Domenico Orlando is
  supported by the European Commission FP6 RTN programme
  MRTN-CT-2004-005104 and in part by the Belgian Federal Science
  Policy Office through the Interuniversity Attraction Pole P5/27 and
  in part by the ``FWO-Vlaanderen'' through project G.0428.06 and
  acknowledges financial support by the Agence Nationale pour la
  Recherche, France, contract 
05-BLAN-0079-01. Marios Petropoulos acknowledges partial financial
  support by the Swiss National Science Foundation.
\end{acknowledgments}

\appendix


\section{The Ricci tensor for the deformed group manifold}
\label{sec:deform-around-wzw}

In order to calculate $R_{\alpha \beta}$, we need to use some
notions in group manifold geometry. Let $\set{\hat \theta^\alpha
}$ be a set of one-forms on a manifold $\mathcal{M}$ satisfying
the commutation relations
\begin{equation}
  \comm{ \hat \theta^\beta , \hat \theta^\gamma  } = \F{\alpha }{\beta \gamma } \hat \theta^\alpha
\end{equation}
as it is the case when $\hat \theta^\alpha $ are the Maurer--Cartan
one-forms for a Lie algebra and $\F{\alpha }{\beta \gamma }$ the
corresponding structure constants. We wish to study the geometry of
the Riemann manifold $\mathcal{M}$ endowed with the metric
\begin{equation}
  g = g_{\alpha \beta } \hat \theta^\alpha  \otimes \hat \theta^\beta  .
\end{equation}
In general such a metric has a symmetry $G \times G^\prime$ where
$G$ is the group corresponding to the structure constants
$\F{\alpha }{\beta \gamma }$ and $G^\prime \subset G$. The
maximally symmetric case, in which $G^ \prime = G$ is obtained
when $g$ is $G$-invariant, \emph{i.e.} when it satisfies
\begin{equation}
  \F{\alpha }{\beta \gamma } g_{\alpha \delta } + \F{\alpha }{\delta \gamma } g_{\alpha \beta }  = 0 .
\end{equation}
For compact groups this condition is fulfilled by the Killing
metric.

The connection one-forms $\omega\ud{\alpha}{\beta}$ are uniquely
determined by the compatibility condition and the vanishing of the
torsion.  Respectively:
\begin{gather}
  \di g_{\alpha \beta } - \omega\ud{\gamma}{\alpha}
  g_{\gamma \beta } - \omega\ud{\gamma}{\beta} g_{\gamma \alpha } = 0, \\
  \di \hat \theta^\alpha + \omega^\alpha_{\phantom{\alpha}\beta }
  \land \hat \theta^\beta = T^\alpha = 0.
\end{gather}
As it is shown in~\cite{Mueller-Hoissen:1988cq}, if $g_{\alpha \beta
}$ is constant, the solution to the system can be put in the form
\begin{equation}
  \omega\ud{\alpha}{\beta}  = -D^{\alpha }_{\hphantom{\alpha}\beta \gamma }\hat \theta^\gamma
\end{equation}
where $D^{\alpha}_{\hphantom{\alpha}\beta \gamma } = 1/2 \F{\alpha
}{\beta \gamma } - K\ud{\alpha}{\beta\gamma }$ and
$K\ud{\alpha}{\beta\gamma }$ is a tensor (symmetric in the lower
indices) given by:
\begin{equation}
  K\ud{\alpha}{\beta \gamma} = \frac{1}{2} g^{\alpha \kappa } \F{\delta }{\kappa \beta } g_{\gamma \delta } +
  \frac{1}{2} g^{\alpha \kappa } \F{\delta }{\kappa \gamma } g_{\beta \delta } .
\end{equation}
The Riemann tensor and the corresponding Ricci tensor now read:
\begin{eqnarray}
  R\ud{\alpha}{\beta \gamma \delta} &=&
  D^\alpha_{\hphantom{\alpha}\beta\kappa} \F{\kappa}{\gamma \delta } +
  D^{\alpha}_{\hphantom{\alpha}\kappa\gamma} D^{\kappa}_{\hphantom{\alpha}\beta\delta} -
  D^{\alpha}_{\hphantom{\alpha}\kappa\delta} D^{\kappa}_{\hphantom{\alpha}\beta\gamma}
\\
  R_{\beta \delta} &=& D^{\alpha}_{\hphantom{\alpha}\beta\kappa}
  \F{\kappa}{\alpha  \delta } - D^{\alpha}_{\hphantom{\alpha}\kappa\delta}
  D^{\kappa}_{\hphantom{\alpha}\beta\alpha}.\label{eq:genric}
\end{eqnarray}

\section{The Jacobi functions}
\label{sec:theta}

We take the standard conventions for the Jacobi functions:
\begin{eqnarray}
  \vartheta_2(z)&=&\sum_{p\in \mathbb{Z}} q^{\frac{1}{2}\left(
  p+\frac{1}{2}\right)^2},\label{theta2} \\
  \vartheta_3(z)&=&\sum_{p\in \mathbb{Z}} q^{\frac{p^2}{2}},
\label{theta3} \\
  \vartheta_4(z)&=&\sum_{p\in \mathbb{Z}} (-1)^p\, q^{\frac{p^2}{2}},
\label{theta4}
\end{eqnarray}
where $q = \exp 2 i \pi z$. With these conventions
\begin{equation}\label{lamellall}
  \lambda = \frac{\vartheta_2^4}{\vartheta_3^4}, \quad
  1- \lambda = \frac{\vartheta_4^4}{\vartheta_3^4}
\end{equation}
and
\begin{equation}\label{eq:dedthet}
2 \eta^3 = \vartheta_2\, \vartheta_3\, \vartheta_4
\end{equation}
with
\begin{equation}\label{eq:ded}
\eta(z) = q^{\frac{1}{24}}\prod_{n=1}^\infty \left(1-q^n \right)
\end{equation}
the Dedekind function. The non-holomorphic function
\begin{equation}\label{eq:E2}
  \hat{E}(z)= \frac{12}{i\pi}\frac{\mathrm{d}\log\eta}{\mathrm{d}z}-\frac{3}{\pi \mathrm{Im\, }z}
\end{equation}
is modular covariant of degree 2.The holomorphic part of the
latter, which is not modular-covariant, is the function $y$
(divided by $i\pi$) in (\ref{eq:parteisencha}).


\bibliography{Biblia}

\end{document}